\documentclass[twocolumn, showpacs, amsmath, amssymb, prl]{revtex4}
\usepackage{graphicx, amssymb, amsmath}
\begin{document}
\title{Granular instability in a vibrated U tube}
\author{I. S\'{a}nchez$^1$, J.R. Darias$^1$, R. Paredes$^2$, C.J. Lobb$^3$, G. Guti\'{e}rrez$^1$}
\affiliation{$^1$Departamento de F\'{\i}sica, Universidad Sim\'{o}n Bol\'{\i}var, Apartado 89000, Caracas 1080-A, Venezuela\\
$^2$Centro de F\'{\i}sica, Instituto Venezolano de Investigaciones Cient\'{\i}ficas, Apartado Postal 21827,
Caracas 1020-A, Venezuela\\
$^3$Department of Physics, University of Maryland, College Park, Maryland 20742-4111, USA}
\date{\today}

\begin{abstract}
We study experimentally the collective motion of grains inside a U shaped tube undergoing vertical oscillations, and we develop a very simple quantitative model that captures relevant features of the observed behaviour. The height difference between the granular columns grows with time when the system is shaken at sufficiently low frequencies. The system exhibits two types of growth: exponentially divergent (type I) and exponentially saturating (type II), depending on the size of the grains. The type I growth can be quenched by removing the air whereas the type II behavior can occur in the absence of air. There is a good agreement between the model proposed and our experimental results.
\end{abstract}
\maketitle

Energy can be injected on a granular system using vibrations to observe many interesting complex phenomena like pattern formation \cite{Aranson}, segregation \cite{Kudrolli}, internal collective motion of the grains, among others \cite{Jaeger}. Tubular containers enhance the role of the walls in the interchange of energy and are extremely important from the industrial point of view. Many industrial processes involve the transport of grains through pipes and are unavoidably affected by mechanical vibrations, therefore their reliability can be greatly affected by the knowledge of the physics of vibrationally induced granular bulk flow \cite{Knowlton}. 

An interesting example of vibrationally induced granular transport in a pipe can be realized in a partially filled U-tube. In this system vertical vibrations can cause the rise of one of the column of grains until the other one is emptied. The U-tube instability was first mentioned by Gutman in 1976 \cite{Gutman}, he pointed out that the increase of the difference in level was due to the air pressure difference acting across the bottom of the vibrated tube. Ohtsuki et al. \cite{Ohtsuki} investigated the change in the level difference for grains in two different size vessels connected at the bottom. They found that friction between particles and the walls of the containers play a key role, and concluded that for large particles the pressure difference due to the interstitial fluid is unimportant.

More recently, other authors have considered the pressure difference due to the interstitial fluid as important for the appearance of the observed instability. Akiyama et al. \cite{Akiyama2} reported measurements of the pressure at the bottom of a partitioned bed linked near the base and vibrated at low frequencies. King et al. \cite{King}, used a vibrated U-tube filled with metal beads and water as an interstitial fluid and measured the pressure drop across the bottom of the container. The experiments reported so far are not conclusive, more research needs to be done to elucidate the mechanisms responsible for these instabilities and many open questions are being investigated [9-12]. Understanding the dynamics of vibrated granular bulk flow in these simple configurations may lead to improvements in the design of sand cores used in metal casting and the transport of granular materials through vertical pipes.

Harmonic vibrations are typically used to fluidize a granular system and a fluidized state can generate flows or hydrostatic like forces. Reyes et al \cite{Reyes}, has shown that the observed fluid like behaviour can be in fact a very complex combination of jammed and air mediated fluidized states that appear during each cycle of oscillation. This cyclic fluidization can produce interesting dynamical features not yet understood, as for example the instabilities reported in this work. To explain our experimental results, we use the idea of cyclic fluidization in a very simple quantitative one dimensional mechanical model. We successfuly describe an exponentially divergent growth (type I) observed in a vibrated U-tube partially filled with grains. A similar model was previously developed to explain reverse buoyancy of a large sphere immersed in a vertically vibrated granular bed \cite{Gutierrez}.
\begin{figure}[t]
\begin{center}
\includegraphics[width=.45\textwidth]{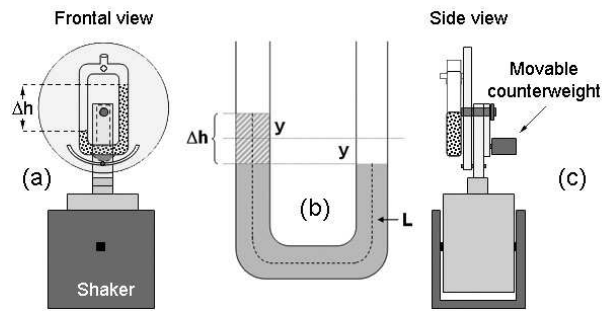}\caption{(a) Sketch of the experimental setup used to study the transport of grains in a vibrated U-tube; (b) Definition of the height difference $\Delta h$, and;(c) Movable counterweight used to align the system and maximize the verticality of the oscillations.}\label{setup}
\end{center}
\end{figure}
In Fig.\ref{setup} we skecth the experimental setup, for the U tube experiments reported here. The tube has inner diameter $D=2$ cm, separation between vertical arms of 6 cm and height of 14 cm. The vertical arms are linked at the top and connected to a vacuum system through two outlets. The tube is fixed to a plexiglas plate with a base attached to an electromagnetic shaker VTS-150. A function generator provided a sinusoidal excitation signal $a_y\sin(\omega t)$ with $\omega =2\pi f$, where $f$ is the frequency, $a_y$ is the vertical amplitude and $t$ is time.

To follow the height difference $\Delta h$ defined in Fig.\ref{setup}(b) between the levels of the granular columns, we used a digital video camera and a stroboscopic light. To achieve maximum verticality in the vibration, we used a counterweight at the back of the plexiglas base, in the form of a pendulum as shown in Fig.\ref{setup}(c). To monitor the verticality of the vibrations we filmed the tube while a stroboscopic light was flashed at a frequency slightly lower than the frequency of vibration of the shaker. Following a dot marked on the vessel we were able to obtain Lissajous figures for the motion of the container. In this way we could conveniently measure the horizontal amplitude $a_x$ and the phase shift $\phi$ between the vertical and horizontal displacements. The grains used were spherical glass beads of different diameters $d$, between $0.18$ mm and $0.36$ mm that we will refer to as small grains, with bulk static bed density $\rho=(1.44\pm 0.04)$ g/cm$^3$ ; and poppy seeds of $d\approx0.8$ mm that we will refer to as large grains, with static bed density $\rho=(0.70\pm 0.05)$ g/cm$^3$. 
\begin{figure}[t]
\begin{center}
\includegraphics[width=.45\textwidth]{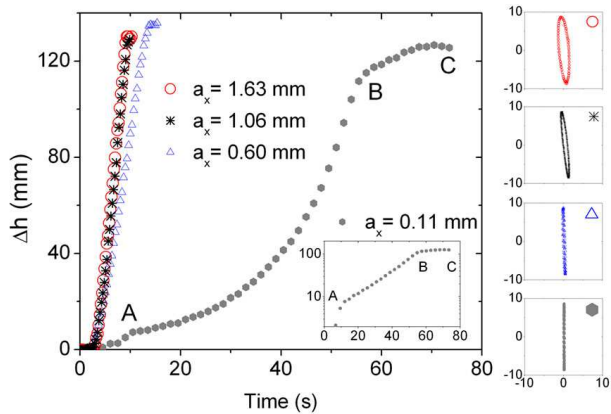}\caption{(Color online) The curve on the right shows the exponential growth of $\Delta h$ with time, for glass spheres with $d=(0.25-0.36)$ mm, vibrated at $f=8$ Hz, and amplitude $a_y=8.5$ mm. The semilog plot in the inset shows the interval where the growth is exponential. Lissajous figures of the motion of the container corresponding to each curve, are shown on the right. The three curves to the left show a linear growth that results from the horizontal component of the driving oscillations.}\label{horizontal}
\end{center}
\end{figure}

Type I growth was found only for small grains, vibrated at low frequencies and specific values of amplitude so that $\frac{a_x}{a_y} \leqslant 0.02$. In Fig.\ref{horizontal} we show the type I growth and the effect of the horizontal component in the behavior of $\Delta h$ as a function of time, for glass beads with $d=(0.25-0.36)$ mm, vibrated at $f=8$ Hz. Different curves are produced changing the position of the movable counterweight shown in Fig.\ref{setup}(c). Here the influence of the horizontal component is clearly shown, the growth of $\Delta h$ changes from exponential to a linear growth as a consequence of the horizontal component of the oscillations. The growth is exponentially divergent in the region between A and B. As the decreasing column becomes empty, the curve deviates from exponential growth and eventually flattens (region B to C). 
\begin{figure}[b]
\begin{center}
\includegraphics[width=.45\textwidth]{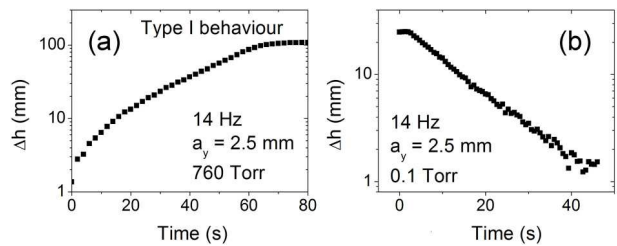}\caption{The role of air is shown for glass spheres with $d=(0.255-0.350)$ mm. (a) Exponentially divergent growth at low frequencies; (b) evacuation of the air up to a pressure of 0.1 Torr, quenches the growth and the columns remain leveled when shaken at low frequency. If they are initially unleveled, $\Delta h$ decays to zero as shown on the graph.}\label{vacuum}
\end{center}
\end{figure}

As mentioned before, interstitial air is necessary for type I growth to occur. When the columns of small grains are set at different levels, and the air is evacuated, $\Delta h$ decays exponentially to zero instead of growing (see Fig. \ref{vacuum}). Air, or more precisely, the insterstitial fluid, plays an important role in many other vertically vibrated granular phenomena [13-20].
\begin{figure}
\begin{center}
\includegraphics[width=.45\textwidth]{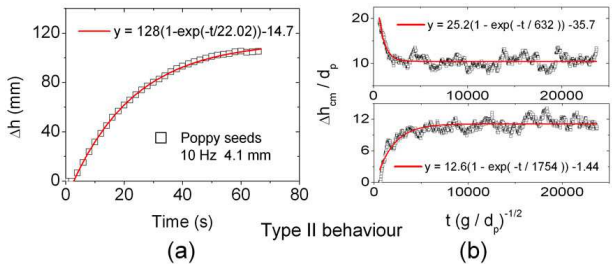}\caption{(Color online) (a) Saturating growth for poppy seeds with $d\approx 0.8$ mm; (b) Simulations showing saturating behaviour for particles with diameter $d_p$ vibrated at $\Gamma = 1.64.$ }\label{saturation}
\end{center}
\end{figure}

When the U tube was partially filled with large grains the vertical vibrations produced a growth of $\Delta h$ that was qualitatively different to the one observed for the small grains. An exponentially saturating growth, that we call type II was obtained for these bigger seeds, as shown in Fig.\ref{saturation}(a). In this case the smaller column does not become empty.

In order to further investigate the behaviour of the large grains in the absence of air and without an horizontal component in the vibration, we performed a molecular dynamics simulations of monodisperse and viscoelastic grains of diameter $d_p=0.8$ mm, with dissipative contact interactions \cite{Ivan}. We used a system of N = 1700 grains in a 2D U-tube of diameter $d = 11.7$ $d_p$, vibrated at $\Gamma = 1.64$, without the presence of air and horizontal component. Results for the simulations are shown in Fig.\ref{saturation}(b). For the lower curve the columns of the granular bed are initially leveled, so that the height difference between the centers of mass of the two columns is $\Delta h_{cm}=0$ $d_p$. The low frequency vibrations produce results similar to those obtained experimentally for poppy seeds. The top curve is produced by same granular bed, under the same conditions, except that initially the two arms are unbalanced so that $\Delta h_{cm}=19$ $d_p$. Due to vibrations, $\Delta h_{cm}$ decreases to the same saturation value of 11 $d_p$. This implies that neither air nor an horizontal component is necessary for the saturating instability to occur.
\begin{figure}[b]
\begin{center}
\includegraphics[width=.45\textwidth]{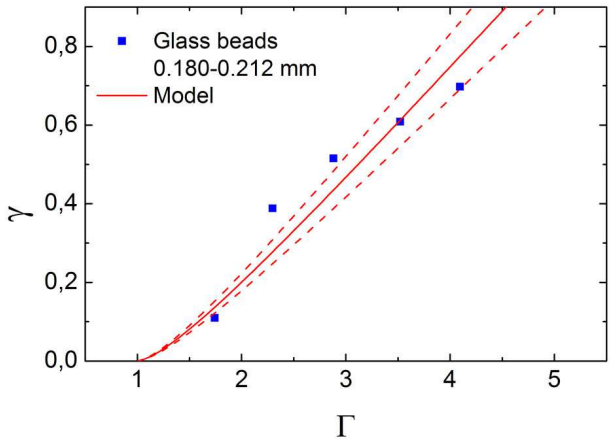}\caption{(Color online) Growth rate, for type I behaviour, as a function of the adimensional acceleration, for glass spheres with $d=(0.18-0.21)$ mm. Three realizations at each value of $\Gamma$ at a fixed frequency $f=12$ Hz, changing the amplitud of the vibration. The solid line represents the model aproximation Eq. (\ref{eq-2}), using the friction coefficient $\nu$ as a free parameter.}\label{gamma&Gamma} 
\end{center}
\end{figure}

We believe that cyclic fluidization is a key mechanism responsible for the growth of $\Delta h$ in a vibrated U tube but there are two qualitatively different processes involved.

A possible qualitative model for the saturating case can be described as follows: the vibrated granular medium acts like a granular bed that is thrown upward, detaches from the container and becomes fluidized, and during flight the bed feels a force due to an effective gravitational field pointing downward, acting as a restoring force. Later, during the settling time, in the same cycle of oscillation, the granular fluid changes to a jammed state by going through a compaction process that causes an increase in the interaction with the walls. This happens in such a way that the largest column ends slightly higher than the shorter one, in each cycle. A quantitative model for this case will be presented in a future work \cite{Ivan}.

The air mediated type I growth, reported here, corresponds to a process where the granular bed, significantly affected by the interstitial air, behaves similarly to a viscous liquid in a U tube, driven by an effective field pulling upward during part of the interval of time when the bed is in a fluidized state. The instability that leads to the type I growth can be described with a simple model similar to the one developed to describe the phenomenon of reverse buoyancy \cite{Gutierrez}. The U tube, partially filled with grains, has cross sectional area $A=\pi D^2/4$. The total mass of the granular bed is $M_g=\rho AL$ where $\rho$ is the static bed density. The vertical position of the tube is given by $w=a_y\cos(\omega t)$.  When the system is fluidized we describe it as a liquid. We assume, for simplicity, that the granular bed fluidizes during an interval of time $\tau$ that coincides with the portion of the period in which the acceleration given by the container is greater than gravity. This assumption could be relaxed to consider the actual time that the system fluidizes. If we define $g_{ef}=a_y\omega^2\cos(\omega t)-g$ as the effective acceleration of the system, then we assume that the system is fluidized while $g_{ef}\geq 0$. This effective acceleration will be approximately equal to the one felt by the granular columns, as long as the bed does not separates significantly from the bottom of the tube. The latter assumption holds for fine grains, at low frequency, in the presence of air. We take $y$ in the growing side of the bed Fig.\ref{setup}(b) as the only degree of freedom necessary to describe the motion of the fluidized bed. In the reference frame of the container, the 1D equation of motion for the evolution of $y(t)$ is given by:
\begin{equation}\label{eq-1}
M_g\ddot y = m(y)g_{ef} - \nu \dot y,
\end{equation}
where $m(y)=2yA\rho$ is the mass of the unbalanced section (striped zone in Fig.\ref{setup}(b)) and we have included a dissipative term to account for friction losses. We integrate this equation over one period of oscillation to get the average displacement $\bar y$, assuming, for this simplified case, that the integral is zero if evaluated outside the interval $[-\tau/2$, $\tau/2]$, where $\tau=\frac{2}{\omega}\cos^{-1}(\Gamma^{-1})$. This gives us a simple differential equation for $\bar y$. If we consider a vertically vibrated U tube with diameter D, partially filled with a granular medium of effective density $\rho$, solving equation (1), we obtain an exponential dependence of $y$ with time, for $\Gamma\ge 1$, with a growing rate $\gamma$ given by:
\begin{equation}\label{eq-2}
\gamma = \frac{g\rho D^2}{2 \nu}\left [\Gamma\sin\left (\cos^{-1}\left (\frac{1}{\Gamma}\right ) \right )-\cos^{-1}\left (\frac{1}{\Gamma}\right )\right ].
\end{equation}
Using the friction coefficient $\nu$ as the only free parameter we obtain a good agreement between experiment and model as can be seen in Fig.\ref{gamma&Gamma}. The dependence on the adimensional acceleration $\Gamma$ of the growing rate $\gamma$ given by equation (2) appears only within the square brackets, but $\rho$ depends on the adimensional acceleration. This dependence can be incorporated by correcting for the expansion of the granular bed as a function of $\Gamma$. Some complications may arise for some configurations because the expansion is not the same in the columns and in the link between them. The friction coefficient may also depend on $\Gamma$. The dotted lines in Fig.\ref{gamma&Gamma} were introduced to account for a 10$\%$ percent variation in the density that occur, when the granular bed is shaken \cite{Ivan}. In our simplified model, we are not taking into account the fact that the actual fluidization time deviates from $\tau$ with increasing $\Gamma$. At higher $\Gamma$ the bed tends to remain fluidized even when $g_{ef} < 0$, lowering $\bar y$ in each cycle, consequently lowering the value of $\gamma$.

The model proposed is crude and could be improved by considering Gutman's model to incorporate the contribution of the pressure difference across the bottom of the tube [5]. It seems plausible that the pressure difference across the bottom contributes to the transfer of grains between the two columns but it is not at all clear how it can act as a force to drive the larger column up in such a way that the growth of $\Delta h$  is exponentially divergent. Although the pressure difference across the bottom grows with the height of the rising granular bed, the mass of the same column also grows linearly with $\Delta h$, and the resultant forces oppose each other. In our model the force due to an effective gravity points upward and increases with increasing mass, producing an exponential growth.  

The fact that the range of parameters for which reverse buoyancy was reported \cite{Gutierrez} (fine grains, small frequencies and large amplitudes) is the same for which exponentially diverging growth was observed, together with the fact that the assumption of cyclic fluidization was successfully used to model both phenomena, gives support to the idea that cyclic fluidization is playing a key role in the dynamics of these interesting instabilities.

We have reported an experimental study of a U-tube partially filled with grains undergoing vertical oscillations. The evolution of the height difference between the levels of the granular material in each side of the tube shows distinct behaviors depending on grain size: type I, wich is an exponentially divergent growth and type II, wich corresponds to a saturating growth. Type II growth can occur in the absence of interstitial fluid, as was confirmed by our computer simulations in a 2D U-tube. The type I growth observed experimentally was modeled assuming cyclic fluidization, and relevant features of the observed behaviour were successfully captured by a very simple quantitative one dimensional model. 

We would like to thank the Decanato de Investigaci\'{o}n of the Universidad Sim\'{o}n Bol\'{\i}var for financial help, Leonardo Reyes for the many helpful discussions, Evelyne Kolb for carefully reading the manuscript, and I. S\'{a}nchez would like to thank Leonardo Ferm\'{\i}n for his contribution in developing the video analysing techniques. This work was sponsored by the FONACIT under Grant S1-2000000624.

\end{document}